\documentclass[traditabstract]{aa}
\usepackage{txfonts}
\usepackage{graphicx}
\usepackage[english]{babel}
\usepackage{mathcomp}
\usepackage{dsfont}
\usepackage{array}

\usepackage{natbib}
\bibpunct{(}{)}{;}{a}{}{,}

\def\be{\begin{equation}}
\def\ee{\end{equation}}
\def\bea{\begin{eqnarray}}
\def\eea{\end{eqnarray}}

\def\fnl{f_\mathrm{NL}}
\def\bps{b_\mathrm{PS}}
\def\air{A_\mathrm{IR}}
\def\bl{b_{\ell_1 \ell_2 \ell_3}}
\def\mbl{\mathbf{b}_{1,2,3}}
\def\lu{\ell_1}
\def\ld{\ell_2}
\def\lt{\ell_3}
\def\dd{\mathrm{d}}
\def\nn{\mathbf{n}}
\def\xx{\mathbf{x}}

\begin{document}

\title{Optimal estimator for the amplitude of the bispectrum from IR sources}  

\author{F. Lacasa\inst{1}\thanks{E-mail: Fabien.Lacasa@ias.u-psud.fr}
\and N. Aghanim\inst{1}\thanks{E-mail: Nabila.Aghanim@ias.u-psud.fr}}

\institute{Institut d'Astrophysique Spatiale (IAS), B\^atiment 121, F-91405 Orsay
(France); Universit\'e Paris-Sud 11 and CNRS (UMR 8617)}

\date{}
 
\abstract
{We devise a fast and optimal estimator for the amplitude of the bispectrum of clustered Infrared (IR) point-sources. We show how this estimator can account for the cases of partial sky coverage and inhomogeneous noise.
Expected detection significance are presented in terms of signal-to-noise, finding that the IR bispectrum will realistically be undetectable below 220 GHz with a Planck-like experiment; on the contrary detection may be achieved at, or above, 220 GHz especially if the CMB is removed. We also show how this estimator can be combined with estimators of radio and CMB non-Gaussianity to build up joint robust constraints. On the one hand, we find that, for a Planck-like experiment, CMB non-Gaussianity estimation can be decoupled from point-source contributions, unless few sources are masked. On the other hand, we find that the estimation of radio and IR non-Gaussianity are strongly coupled, which diminishes their separate detection significance.
}


\maketitle

%******************************************************************

\section{Introduction}

The measurement and study of high order moments has emerged as an important field in observational cosmology as it probes deviations from Gaussianity. For example, observing non-Gaussianity (NG) from the Cosmic Microwave Background (CMB) or the large scale structure would probe primordial NG dating back to the generation of cosmological perturbations ; this in turn would constrain severely primordial models generating these fluctuations (e.g. inflation). For example, a NG detection with the sensitivities of today's experiments would rule out a whole class of inflation models, among which the standard single-field slow roll inflation \citep{Acquaviva2003,Maldacena2003,Creminelli2004}.

The most popular form of primordial non-Gaussianity is the so-called `local' type, parametrised by a factor $\fnl$, for which the Bardeen potential takes the form :
\be
\Phi(\xx) = \Phi_G(\xx) + \fnl \left( \Phi_G(\xx)^2 - \langle  \Phi_G(\xx)^2 \rangle \right)
\ee
where $\Phi_G$ is the Gaussian part of the potential. It is on this form of NG that most efforts have focused and that the tightest constraints were obtained. Indeed a fast estimators for $\fnl$ was developed and applied to the CMB data \citep{KSW2005,Komatsu2009,Komatsu2011}.
However at CMB frequencies, many contaminating signals are present, which typically have a non-Gaussian distribution. While the galactic emission is mainly confined to a defined area : the galactic plane, and may thus be masked, extragalactic point-sources are present all over the sky and the faintest (below detection limits and confusion levels) are neither detected nor masked. These unresolved sources are therefore one potentially important source of secondary non-Gaussianity that needs to be estimated.

There are principally two populations of point-sources of interest for the CMB : radio-loud sources with strong magnetic fields yielding synchrotron or free-free emission in the radio domain, and Dusty Star-Forming Galaxies (DSFGs) with thermal emission from dust heated by young star in the infrared (IR) domain. Radio galaxies can be considered randomly distributed on the sky \citep{Toffolatti1998,Gonzalez-Nuevo2005} at the CMB frequencies and thus have flat moments. Their skewness is related to their number counts as~:
\be
\bps = \int S^3 \, \frac{\dd n}{\dd S} \, \dd S
\ee
where $S$ is the source flux in Jy ($\bps$ may then be converted to CMB temperature elevation through Planck's law, but is usually quoted in dimensionless units $\Delta T/T$ in the literature).\\
On the contrary, infrared galaxies are strongly clustered in dark matter halos, which typically enhances their correlation functions on large scales. Most recent CMB experiments, e.g. Planck \citep{PlanckBluebook}, ACT \citep{Das2011} and SPT \citep{Keisler2011}, are probing higher frequencies compared to previous missions so that the modeling and measurement of the infrared background becomes important. The Cosmic Infrared Background (CIB) and its anisotropies have indeed been detected and characterised at the power spectrum level at these frequencies \citep[e.g.][]{Hall2010, Planck-early-CIB, Viero2012}.

Non-Gaussianity arising from point sources (radio or IR) is of particular importance for CMB studies, where maps are contaminated by foregrounds which are imperfectly subtracted by component separation methods. Therefore estimation of the primordial NG parameter $\fnl$ may be contaminated by the point-source NG \citep{Babich2008,Lacasa2012}. Measurement of the non-Gaussianity of point-sources could also be a new way to further constrain the respective models for their emission and/or spatial distribution. 

A fast estimators for $\bps$ has been developed and applied to CMB data \citep{KSW2005,Komatsu2009,Komatsu2011}. Taking advantage of a prescription for the non-Gaussianity of IR source developed by \citep{Lacasa2012}, the goal of this article is to propose a fast estimator for the amplitude of the IR bispectrum. This can be used in non-Gaussianity studies to disentangle the primordial and foreground contributions so that measurements of both of them are unbiased.

The article is organised as follows : in section \ref{Sect:IRbisp} we briefly recall the prescription for the bispectrum of IR sources and introduce the estimator for its amplitude,  in section \ref{Sect:maskandnoise} we tackle the problems of partial sky coverage and anisotropic noise,  in section \ref{Sect:results} we present the expected signal-to-noise ratio (SNR) of detection at several frequencies, in section \ref{Sect:jointNG} we show how to combine this estimator with the ones used for the CMB and radio galaxies to produce a joint constraint of non-Gaussianity, and we conclude in section \ref{Sect:concl}.

%**************************************************************************

\section{Bispectrum of IR sources}\label{Sect:IRbisp}
\subsection{Bispectrum}
Given a full-sky map of the temperature fluctuations $\Delta
T(\nn)$ of some signal in direction $\nn$, it can be decomposed in the spherical
harmonic basis 
\begin{equation}
a_{\ell m} = \int \dd^2\nn \; Y^*_{\ell m}(\nn) \; \Delta
T(\nn)
\end{equation}
with the usual orthonormal spherical harmonics $Y_{\ell m}$
\begin{equation}
\int \dd^2 \nn \; Y_{\ell m}(\nn) \; Y^*_{\ell'
  m'}(\nn) \, 
= \, \delta_{\ell \ell'} \, \delta_{m m'}.
\end{equation}
The power spectrum $C_\ell$ of the signal is the 2-point correlation function in harmonic space given as :
\be
\langle a_{\ell m} \, a^*_{\ell' m'} \rangle = C_\ell \, \delta_{\ell \ell'} \, \delta_{m m'}
\ee
The bispectrum $\bl$ is the 3-point correlation function in harmonic space:
\be\label{Eq:defbisp}
\langle a_{\ell_1 m_1} a_{\ell_2 m_2} a_{\ell_3 m_3} \rangle = G_{\lu \ld \lt}^{m_1 m_2 m_3} \times \bl
\ee
with the Gaunt coefficient
\bea
G_{1,2,3} &=& \int \dd^2\nn \,Y_{123}(\nn) \\
&=& \sqrt{\frac{(2\ell+1)_{123}}{4\pi}} 
\left(\begin{array}{ccc}
\ell_1 & \ell_2 & \ell_3 \\ 0 & 0 & 0
\end{array}\right)
\left(\begin{array}{ccc}
  \ell_1 & \ell_2 & \ell_3\\
  m_1 & m_2 & m_3
\end{array}\right)
\eea
where $Y_i=Y_{\ell_i m_i}$. In the following the subscript 123 denotes the product of the corresponding variables e.g. $X_{123} \equiv X_1 \, X_2 \, X_3$. $G_{1,2,3}$ is zero unless the triplet $(\ell_1,\ell_2,\ell_3)$ follows the triangle inequalities and $m_1+m_2+m_3=0$.\\
The bispectrum estimator is :
\be
\hat{b}_{123} = \frac{1}{N_{123}} \sum_{m_1,m_2,m_3} G_{1,2,3}  \; a_1 \, a_2 \, a_3
\ee
with
\be
N_{\lu \ld \lt} = \frac{(2\ell+1)_{123}}{4\pi}
\left(\begin{array}{ccc}
\ell_1 & \ell_2 & \ell_3 \\ 0 & 0 & 0
\end{array}\right)^2 
\ee
being the `number of modes' for the $(\lu,\ld,\lt)$ triangle.

In the following, we note in bold the bispectrum which accounts for beam effect~:
$$\mbl = \bl \; b_{\ell_1} b_{\ell_2} b_{\ell_3}$$.
We also note
$${\cal C}_\ell = C_\ell\; b^2_\ell + C_\ell^\mathrm{noise}$$
The power spectrum accounting for both the noise and the beam effects.

\subsection{Optimal estimator of the amplitude of the bispectrum}
The non-Gaussianity from radio sources at CMB frequencies is entirely characterised by a single parameter $\bps$ which can be estimated on a map \citep{KSW2005}. Hence the radio-source contamination to $\fnl$ can be marginalised over. By contrast, the IR non-Gaussianity is more complicated and is characterised by the full bispectrum, whose estimation is computationally intensive (with $\mathcal{O}(N_\mathrm{pix}^{5/2})$ operations). 

In this context, \cite{Lacasa2012} have proposed a full-sky analytical prescription to compute the bispectrum of extra-galactic sources from their number counts and power spectra. In this prescription, the bispectrum arising from a population of point sources writes :
\be\label{Eq:presp}
\bl^\mathrm{PS} = \alpha \sqrt{C_{\lu} C_{\ld} C_{\lt}}
\ee
with $C_\ell$ being the power spectrum of the considered point sources, and
\be
\alpha = \frac{\int S^3 \frac{\dd N}{\dd S} \dd S}{\left(\int S^2 \frac{\dd N}{\dd S} \dd S\right)^{3/2}}
\ee
where $S$ is the source flux, the integral runs up to the detection limit of the resolved sources indicated by a flux cut $S_\mathrm{cut}$, and $\frac{\dd N}{\dd S}$ are the number counts per steradian.\\
 This prescription gives a bispectrum in good agreement with the one measured on maps of radio and dusty star-forming galaxies simulated by \cite{Sehgal2010}.\\
In the following, we will focus solely on IR point-sources. The level of non-Gaussianity induced by IR sources can thus be quantified by the amplitude of their bispectrum compared to the template bispectrum given by Eq.\ref{Eq:presp}. The use of the amplitude of the bispectrum permits to derive easily the bias on primordial CMB estimation $\fnl$, and to marginalise over this nuisance parameter as will be seen in Sect.\ref{Sect:jointNG}.

The amplitude of the bispectrum is measured by minimising the $\chi^2$ of the map bispectrum to the theoretical one :
\be
\chi^2(A) = \sum_{\lu \leq \ld \leq \lt} \frac{\left(\mbl - \mathrm{A}\,\mbl^\mathrm{IR}\right)^2}{\sigma^2(\lu,\ld,\lt)}
\ee
with
\be \label{Eq:defsig2l123}
\sigma^2(\lu,\ld,\lt) = \frac{{\cal C}_{\lu}^\mathrm{tot} {\cal C}_{\ld}^\mathrm{tot} {\cal C}_{\lt}^\mathrm{tot}}{N_{\lu \ld \lt}}\times \Delta_{\lu \ld \lt}
\ee
being the bispectrum variance in the weak non-Gaussianity limit\footnote{This approximation is justified for the Cosmic Infrared Background (CIB) which is close to Gaussian (see \citep{Lacasa2012} for details)}.
In this expression, ${\cal C}_{\ell}^\mathrm{tot}$ is the map power spectrum, and we have
\bea
\nonumber\Delta_{\lu \ld \lt} &=& 1+ \delta_{\lu\ld}+ \delta_{\lu\lt}+ \delta_{\ld\lt}+ 2\,\delta_{\lu\ld\lt}\\
&=& \left\{ \begin{array}{ll} 6 & \mathrm{equilateral \; triangle}\\ 2 &
  \mathrm{isosceles \; triangle}\\ 1 & \mathrm{general
    \; triangle} \end{array}\right.
\eea
Then the maximum likelihood estimator is:
\be\label{Eq:hatA}
\hat{\mathrm{A}} = \sum_{\lu \leq \ld \leq \lt} \frac{\mbl \;\mbl^\mathrm{IR}}{\sigma^2(\lu,\ld,\lt)} \times \sigma^{-2}(\hat{\mathrm{A}})
\ee
with 
\be
\sigma^2(\hat{\mathrm{A}}) = \sum_{\lu \leq \ld \leq \lt} \frac{\left(\mbl^\mathrm{IR}\right)^2}{\sigma^2(\lu,\ld,\lt)}
\ee
being a normalisation factor giving the error-bar in the assumption of a Gaussian likelihood, which is valid in the weak NG limit when $\sigma^2(\lu,\ld,\lt)$ can be considered independent of A.\\
The theoretical bispectrum, Eq.\ref{Eq:presp}, is separable\footnote{a bispectrum is said separable if it can be written in the form $b(\ell_1,\ell_2,\ell_3) = \sum_i f_i(\ell_1) \, g_i(\ell_2) \, h_i(\ell_3) + \mathrm{perm.}$ This is the case for the IR bispectrum with the sum reducing to 1 element and $f_i(\ell) = g_i(\ell) = h_i(\ell) = \alpha^{1/3} \sqrt{C_\ell^\mathrm{IR}}$} ; hence a simpler computation of the numerator of Eq.\ref{Eq:hatA} can be devised, inspired by \cite{KSW2005}. Indeed, defining the filtered map:
\be\label{Eq:filteredF}
F(\nn) = \sum_{\ell m} \frac{\alpha^{1/3} \, \sqrt{C_\ell^{\mathrm{IR}}} \, b_\ell }{{\cal C}_\ell^\mathrm{tot}}\, a_{\ell m} \, Y_{\ell m}(\nn)
\ee
and
\be
\mathcal{S}_\mathrm{IR}= \int \dd^2\nn \; F(\nn)^3
\ee
we have:
\be \label{Eq:defSir}
\mathcal{S}_\mathrm{IR}= \sum_{\lu \ld \lt} \frac{\mbl^\mathrm{IR} \; \hat{\mathbf{b}}_{123}}{{\cal C}_{\lu}^\mathrm{tot} {\cal C}_{\ld}^\mathrm{tot} {\cal C}_{\lt}^\mathrm{tot}} \times N_{123} = 6 \times \sum_{\lu \leq \ld \leq \lt} \frac{\mbl^\mathrm{IR} \; \hat{\mathbf{b}}_{123}}{\sigma^2(\lu,\ld,\lt)}
\ee
Furthermore it is worth noting that this approach is faster than a full bispectrum analysis, having $\mathcal{O}(N_\mathrm{pix}^{3/2})$ operations. 

%**************************************************************************

\section{Masked sky and inhomogeneous noise}\label{Sect:maskandnoise}
In realistic cases of CMB analysis, the statistical isotropy of the CMB signal is broken by inhomogeneous noise --e.g. due to scanning strategy-- or by masking large areas of the sky --e.g. those contaminated by galactic emissions. In these cases, the estimator of the IR-bispectrum amplitude is no longer optimal and it is biased in a non-trivial way. Nevertheless and similarly to the case of the $f_{\mathrm{NL}}$, the bias and the lack of optimality can both be tackled through adapted modifications to the estimator.

\subsection{Debiasing}\label{Sect:debias}

When a covariance matrix of the map(s) can be estimated, \cite{Creminelli2006} have shown that Wiener filtering of the map(s) will debias the non-Gaussian estimator from anisotropic contaminants or noise. Specifically, if $\mathbf{C}$ is the estimated covariance matrix in harmonic space, the estimator Eq. \ref{Eq:hatA} can be debiased by making the change
\be
a_{\ell m} \rightarrow \left(\mathbf{C}^{-1}\cdot \mathbf{a}\right)_{\ell m} = \sum_{\ell' m'} C^{-1}_{\ell m, \ell' m'} \; a_{\ell' m'}
\ee
in the filtered map Eq. \ref{Eq:filteredF}, where $\mathbf{a} = \left(a_{\ell m}\right)_{\ell=2..\ell_\mathrm{max}\, , \, m=-\ell..\ell}$ is the vector of harmonic coefficients.

This step is sometimes called $\mathbf{C}^{-1}$ prefiltering, and replaces the $1/{\cal C}_\ell^\mathrm{tot}$ filtering of Eq.\ref{Eq:filteredF} (in the isotropic case $\mathbf{C}^{-1}$ reduces to $1/{\cal C}_\ell^\mathrm{tot}$). An efficient algorithm for this filtering has been designed by \cite{Elsner2012}.

\subsection{Reducing variance and reaching optimality} \label{Sect:optimalisation}
When isotropy is broken, the 3-point correlation function used to define the bispectrum in Eq. \ref{Eq:defbisp} no longer has minimal variance. It must be replaced by the Wick product of the three harmonic coefficients given by:
\bea
\nonumber a_{\ell_1 m_1} \, a_{\ell_2 m_2} \, a_{\ell_3 m_3} &\rightarrow& a_{\ell_1 m_1} \, a_{\ell_2 m_2} \, a_{\ell_3 m_3} - \langle a_{\ell_1 m_1} \, a_{\ell_2 m_2} \rangle \, a_{\ell_3 m_3} \\
&&- \langle a_{\ell_1 m_1} \, a_{\ell_3 m_3} \rangle \, a_{\ell_2 m_2} - \langle a_{\ell_2 m_2} \, a_{\ell_3 m_3} \rangle \, a_{\ell_1 m_1}.
\label{eq:lin}
\eea
This expression exhibits the same mean as in the isotropic case but it has a lower variance. 
Furthermore when applied to the isotropic case Eq. \ref{eq:lin} gives the same bispectrum estimator as in Eq. \ref{Eq:defbisp} (e.g. for $\lu\neq\ld\neq\lt$ all the expectation values vanish).\\ Only the 2-point correlation function is to be considered for the expectation values. Therefore in practice, the latter are obtained from a sufficiently large number of Gaussian realisations with same power spectrum as that of the signal considered.\\
Inputting the linear corrections given by Eq. \ref{eq:lin} into the expression of $S_\mathrm{IR}$ in Eq. \ref{Eq:defSir} yields:
\be
S_\mathrm{IR} \rightarrow \tilde{S}_\mathrm{IR} = \int \dd^2\nn \; F(\nn)^3 -3\times \!\!\int \dd^2\nn \; F(\nn) \, G(\nn)
\ee
with
\bea
G(\nn) &=& \sum_{12} \alpha^{2/3} \sqrt{C_{\lu}^{IR} \, C_{\ld}^{IR}} \; \langle a_1 \, a_2 \rangle_\mathrm{MC}\; Y_1(\nn)\, Y_2(\nn)\\
&=& \langle F(\nn)^2 \rangle_\mathrm{MC}
\eea

%**************************************************************************

\section{Expected detection significance}\label{Sect:results}

In this section, we present the expected detection significance in terms of the signal-to-noise ratio for the previously described estimator. To this end, we use values of $\alpha$ and $C_\ell^\mathrm{IR}$ that we measure in the simulations of \cite{Sehgal2010} at frequencies 150, 220, 280 and 350 GHz. The obtained values were shown to reproduce adequately the IR bispectrum at these frequencies (see \citet{Lacasa2012}).

\subsection{Ideal cosmic variance-limited case} 

Let us first focus on an ideal case where the signal is made solely of IR galaxies, i.e. noise and other astrophysical contributions are neglected. We will compute the signal-to-noise ratio of the detection of the non-Gaussian signal as defined by the amplitude of the bispectrum. Each triplet $\lu\leq\ld\leq\lt$ contains information proportional to its number of configurations :
\be
\sigma^2(\hat{\mathrm{A}}) = \alpha^2 \sum_{\lu \leq \ld \leq \lt} N_{\lu \ld \lt}
\ee
In other terms, the SNR of the detection goes down with the total number of bispectrum configurations:
\be
\mathrm{SNR} = \frac{1}{\sigma(\hat{\mathrm{A}})} \propto 1/\sqrt{N_\mathrm{tot}(\ell_\mathrm{max})}
\ee
with
\be
N_\mathrm{tot}(\ell_\mathrm{max}) = \!\!\!\!\!\!\!\!\!\sum_{\ell_\mathrm{min}\leq\lu \leq \ld \leq \lt\leq\ell_\mathrm{max}}\!\!\! N_{\lu \ld \lt}
\ee
For example in the case of full-sky maps of the IR sources at frequencies 150, 220, 280 and 350~GHz with angular scales ranging from $\ell_\mathrm{min}=2$ to $\ell_\mathrm{max}=2048$, the obtained SNRs are of order $\sim$1000 and they do not vary much from frequency to frequency (see first line of Table \ref{Table:SNR}).

\subsection{Realistic case}
In a real experiment, the IR signal/map is convolved by the instrumental beam and the signal is contaminated by noise and other astrophysical signals, the most prominent being the CMB and the galaxy, which can be masked. Hence the SNR is significantly decreased compared to the ideal case. Table \ref{Table:SNR} summarizes the expected SNR in the case of a 5 arcmin beam and $10^{-8}$ dT/T homogeneous detector noise across frequencies. We show how the detection level/significance decrease from the ideal case to the case of a convolved but "perfect" full-sky IR map, then including CMB contamination but still full-sky, and finally with a 50\% sky-fraction mimicking the mask that may be applied to avoid galactic dust emission. We also include the case where 90\% of the CMB can be removed by a component separation method. For the partial-sky coverage cases, we assume that the optimisation and the debiasing described in Sect. \ref{Sect:debias}\&\ref{Sect:optimalisation} have been applied with the SNR scaling as $f_\mathrm{SKY}^{-1/2}$.

\begin{table}[ht!]
\centering
\begin{tabular}{|c|c|c|c|c|}
\hline
frequency (GHz) & 150 & 220 & 280 & 350 \\
\hline
Ideal case & 1218 & 1157 & 1161 & 1159 \\
Full-sky IR & 15.5 & 98 & 336 & 833 \\
Full-sky with CMB & 0.39 & 6.7 & 55 & 387 \\
50\% sky with CMB & 0.28 & 4.7 & 39 & 274 \\
50\% sky with 10\% CMB & 3.67 & 45.9 & 210 & 577 \\
\hline
\end{tabular}
\caption{Expected SNR at $\ell_\mathrm{max}=2048$. Ideal stands for a full-sky cosmic-variance limited IR map without noise nor contaminations. Full-sky IR represents the case where the IR map is convolved with a 5 arcmin Gaussian beam and contains $10^{-8}$ $\Delta T/T$ noise ($C_\ell = 7.4 \cdot10^{-4} \mu K^2\cdot\mathrm{sr}$). The same instrumental specifications are adopted for the following lines. Full-sky with CMB is when CMB is present, contaminating the IR map. 50\% sky with CMB stands for the case of masking half of the sky. 50\% sky with 10\% CMB stands for a half sky IR map where 90\% of the CMB (in amplitude) has been removed by some component separation.
}
\label{Table:SNR}
\end{table}

The SNR of the detection is severely reduced from ideal to more realistic cases especially at lower frequencies. The signal is still detectable at all frequencies when including beam and noise effects. However, when the CMB is included the signal becomes undetectable at 150 GHz. Above this frequency the detection significance is SNR$\sim$5$\sigma$ at 220 GHz and reaches 40 to 200$\sigma$ at 280 and 350 GHz where the IR emission is dominant even in presence of CMB contribution. \\
Removing the CMB, with a component separation method, permits to further improve the detection at 220 GHz by increasing the SNR to $\sim$46$\sigma$. In this case, the IR non-Gaussianity at 150 GHz seems marginally detectable. However at this frequency the non-Gaussianity from radio sources becomes an issue (while at higher frequencies IR sources are forecasted to dominate the unresolved sources population). We tackle in the next section the issue of non-Gaussianity other than that of the IR sources.

%**************************************************************************

\section{Joint NG estimation}\label{Sect:jointNG}

When several non-Gaussian signals are present, a joint estimation of their amplitudes taking their covariances into account is necessary. We focus in the present study on the main extragalactic non-Gaussian signals, namely CMB, radio and IR point-sources. A joint estimation requires minimisation of the chi-square :
\be\label{Eq:chi2fnlbpsAIR}
\chi^2(\fnl, \bps, \air) = \sum_{\lu \leq \ld \leq \lt} \frac{\left(\mbl^\mathrm{obs} - \mbl^\mathrm{model}(\fnl, \bps, \air)\right)^2}{\sigma^2(\lu,\ld,\lt)}
\ee
with
\be
\mbl^\mathrm{model}(\fnl, \bps, \air) = 
\fnl \,\mbl^\mathrm{CMB} + \bps \,\mbl^\mathrm{RAD} + \mathrm{A}_\mathrm{IR}\,\mbl^\mathrm{IR}
\ee
and (involved in the definition of $\sigma^2(\lu,\ld,\lt)$, see Eq.\ref{Eq:defsig2l123})
\be
{\cal C}_\ell^\mathrm{tot} = \left(C_\ell^\mathrm{CMB}+C_\ell^\mathrm{RAD}+C_\ell^\mathrm{IR}\right) b_\ell^2 + C_\ell^\mathrm{noise}
\ee
We use the standard' parameter $\bps$ for compatibility with the literature (instead of noting e.g. $A_\mathrm{RAD}$).\\
Then if we define the scalar product between two bispectra $\mathbf{b}^\alpha$ and $\mathbf{b}^\beta$ :
\be\label{Eq:defscalprod}
< \mathbf{b}^\alpha , \mathbf{b}^\beta > = \sum_{\lu \leq \ld \leq \lt} \frac{\mbl^\alpha \mbl^\alpha}{\sigma^2(\lu,\ld,\lt)}
\ee
Minimising Eq.\ref{Eq:chi2fnlbpsAIR} corresponds to solving the linear system :
\bea \label{Eq:JointNG_scalprod}
\nonumber &\left(\begin{array}{ccc}
<\mathbf{b}_\mathrm{CMB} , \mathbf{b}_\mathrm{CMB}> & <\mathbf{b}_\mathrm{RAD} , \mathbf{b}_\mathrm{CMB}> & <\mathbf{b}_\mathrm{IR} , \mathbf{b}_\mathrm{CMB}> \\
<\mathbf{b}_\mathrm{CMB} , \mathbf{b}_\mathrm{RAD}> & <\mathbf{b}_\mathrm{RAD} , \mathbf{b}_\mathrm{RAD}> & <\mathbf{b}_\mathrm{IR} , \mathbf{b}_\mathrm{RAD}> \\
<\mathbf{b}_\mathrm{CMB} , \mathbf{b}_\mathrm{IR}> & <\mathbf{b}_\mathrm{RAD} , \mathbf{b}_\mathrm{IR}> & <\mathbf{b}_\mathrm{IR} , \mathbf{b}_\mathrm{IR}>
\end{array}\right)
\cdot
\left(\begin{array}{c}
\fnl \\
\bps \\
\air
\end{array}\right)
\\ &\quad =
\left(\begin{array}{c}
<\mathbf{b}_\mathrm{obs} , \mathbf{b}_\mathrm{CMB}> \\
<\mathbf{b}_\mathrm{obs} , \mathbf{b}_\mathrm{RAD}> \\
<\mathbf{b}_\mathrm{obs} , \mathbf{b}_\mathrm{IR}>
\end{array}\right)
\eea
If we introduce --noted with upper tilde-- the ``naive" estimators which consider only one source of non-Gaussianity, e.g.
\be
\tilde{A}_\mathrm{IR} = \frac{<\mathbf{b}_\mathrm{obs} , \mathbf{b}_\mathrm{IR}>}{<\mathbf{b}_\mathrm{IR} , \mathbf{b}_\mathrm{IR}>}
\ee
then Eq.\ref{Eq:JointNG_scalprod} can be rewritten to define the Joint NG estimators --noted with upper hat-- as :
\be
\left(\begin{array}{c}
\hat{f}_\mathrm{NL} \\
\hat{b}_\mathrm{PS} \\
\hat{A}_\mathrm{IR}
\end{array}\right)
=
\mathcal{M}^{-1} \cdot
\left(\begin{array}{c}
\tilde{f}_\mathrm{NL} \\
\tilde{b}_\mathrm{PS} \\
\tilde{A}_\mathrm{IR}
\end{array}\right)
\ee
with
\be
\mathcal{M} = \left(\begin{array}{ccc}
1 & \Delta \fnl^\mathrm{RAD} &  \Delta \fnl^\mathrm{IR}\\
\Delta \bps^\mathrm{CMB} & 1 & \Delta \bps^\mathrm{IR} \\
\Delta \air^\mathrm{CMB} & \Delta \air^\mathrm{RAD} & 1
\end{array}\right)
\ee
and
\bea
\nonumber \Delta \fnl^\mathrm{RAD} = \frac{<\mathbf{b}_\mathrm{RAD} , \mathbf{b}_\mathrm{CMB}>}{<\mathbf{b}_\mathrm{CMB} , \mathbf{b}_\mathrm{CMB}>} \qquad
\Delta \fnl^\mathrm{IR} =\frac{<\mathbf{b}_\mathrm{IR} , \mathbf{b}_\mathrm{CMB}>}{<\mathbf{b}_\mathrm{CMB} , \mathbf{b}_\mathrm{CMB}>}\\
\nonumber \Delta \bps^\mathrm{CMB} = \frac{<\mathbf{b}_\mathrm{CMB} , \mathbf{b}_\mathrm{RAD}>}{<\mathbf{b}_\mathrm{RAD} , \mathbf{b}_\mathrm{RAD}>} \qquad
\Delta \bps^\mathrm{IR} = \frac{<\mathbf{b}_\mathrm{IR} , \mathbf{b}_\mathrm{RAD}>}{<\mathbf{b}_\mathrm{RAD} , \mathbf{b}_\mathrm{RAD}>}\\
\nonumber \Delta \air^\mathrm{CMB} = \frac{<\mathbf{b}_\mathrm{CMB} , \mathbf{b}_\mathrm{IR}>}{<\mathbf{b}_\mathrm{IR} , \mathbf{b}_\mathrm{IR}>} \qquad
\Delta \air^\mathrm{RAD} = \frac{<\mathbf{b}_\mathrm{RAD} , \mathbf{b}_\mathrm{IR}>}{<\mathbf{b}_\mathrm{IR} , \mathbf{b}_\mathrm{IR}>}
\eea

We compute for illustration the mixing matrix at 220 GHz. Neglecting the noise and beam effect and assuming the ERCSC flux cuts \citep{Planck-ERCSC}, we find for $\ell_\mathrm{max}=2048$ :
\be
\mathcal{M} = \left(\begin{array}{ccc}
1 & 0.124 &  0.631 \\
2.08 \!\cdot\! 10^{-5} & 1 & 1.79 \\
3.16 \!\cdot\! 10^{-5} & 0.535 & 1
\end{array}\right)
\ee
where we renormalised the radio amplitude to get numbers of order 1, i.e., the mixing matrix must be understood to be applied to $\bps/(1.8\!\cdot\! 10^{-28})$.\\
The matrix is mostly bloc diagonal, with the $\fnl$ part decoupling from the $\bps$ and $\air$ parts. Therefore, unless $\bps \gg 10^{-28}$, $\fnl$ estimation can be considered unaffected by extragalactic foregrounds at this frequency. Conversely, estimation of NG from radio and IR is unaffected by $\fnl$, but however we see that the two contributions affect heavily each other, with the $\bps$-$\air$ submatrix being poorly conditioned (condition number $\sim 180$).

When we include instrumental effects, namely a 5 arcmin Gaussian beam and a $10^{-8} \Delta T/T$ noise (same $\ell_\mathrm{max}$ and flux cut), we find the mixing matrix :
\be
\mathcal{M} = \left(\begin{array}{ccc}
1 & 0.0461 &  0.245 \\
1.21\!\cdot\! 10^{-4} & 1 & 2.12 \\
1.31\!\cdot\! 10^{-4} & 0.435 & 1
\end{array}\right)
\ee
The relative importance of CMB NG compared to point-sources NG has increased. Indeed beam and noise effects mostly decrease the signal-to-noise in configurations with high multipoles ; this down-weights point-sources NG which is dominant at high $\ell$, while CMB NG dominates at lower multipoles. However CMB and point-sources NG estimation can still be considered decoupled unless $\bps \gg 10^{-28}$. In the $\bps$-$\air$ submatrix, we see that the relative importance of $\bps$ has decreased compared to $\air$. Indeed the IR signal is most important at large angular scales while the radio SNR comes mostly from small angular scales which are affected by the beam smoothing and the noise.

The $\bps$-$\air$ submatrix is still ill-conditioned when we account for beam and noise effect ; this is due to a high correlation between the IR and radio bispectra templates. This correlation can be quantified by the quantity $r = \cos \theta \in [-1,1]$ :
\be
r_{\alpha,\beta} = \frac{<\mathbf{b}_\alpha , \mathbf{b}_\beta>}{\sqrt{<\mathbf{b}_\alpha , \mathbf{b}_\alpha> \, <\mathbf{b}_\beta , \mathbf{b}_\beta>}}
\ee
with $\alpha,\beta=$ RAD,IR or CMB. In other terms $\theta_{\alpha,\beta}$ is the angle between the vectors $\mathbf{b}_\alpha , \mathbf{b}_\beta$ in the vector space of bispectra with the scalar product defined in Eq.\ref{Eq:defscalprod}. At $\ell_\mathrm{max} = 2048$, with a 5 arcmin Gaussian beam and $10^{-8} \Delta T/T$ noise, the correlation coefficients of the different bispectra are listed in Table \ref{Table:correlcoef}. 

\begin{table}[ht!]
\centering
\begin{tabular}{|c|c|c|c|}
\hline
 & IR/RAD & IR/CMB & RAD/CMB \\
\hline
r & 96\% & 0.57\% & 0.24\% \\
\hline
% ideal & 98\% & 0.45\% & 0.16\% \\
%\hline
\end{tabular}
\caption{Correlation coefficients between the different bispectrum templates at 220 GHz with $\ell_\mathrm{max} = 2048$, 5 arcmin Gaussian beam and $10^{-8} \Delta T/T$ noise.}
\label{Table:correlcoef}
\end{table}

\begin{figure}[htbp]
\begin{center}
\includegraphics[width=1.\linewidth]{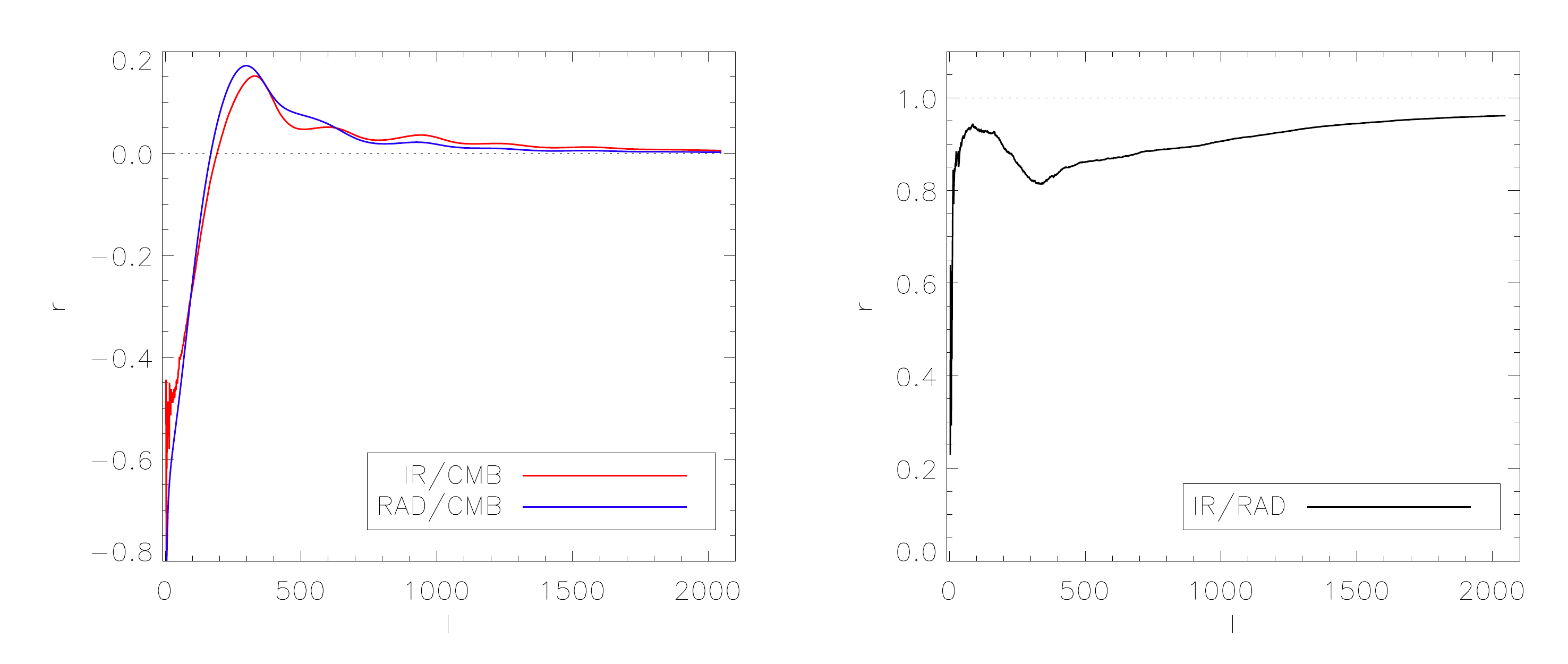}
\caption{Correlation coefficients between the radio, IR and CMB bispectra at 220 GHz with 5 arcmin Gaussian beam and $10^{-8}\, \Delta T/T$ noise. Left panel : correlation between IR and CMB in red, between radio and CMB in blue, as a function of $\ell_{\mathrm{max}}$. Right panel : correlation between IR and radio.}
\label{Fig:correlcoef}
\end{center}
\end{figure}

In Fig.\ref{Fig:correlcoef}, we plot these coefficients as a function of the maximum multipole used in the analysis, with the same instrumental specifications.

When analysing only large angular scales (small $\ell_{\mathrm{max}}$), the radio and IR bispectrum templates are anti-correlated with the CMB bispectrum. This is because both radio and IR bispectra estimators are positive while the CMB is negative due to the Sachs-Wolfe effect term. However when going to higher resolution, the first acoustic peak makes the CMB bispectrum positive with large values in some configurations (e.g. equilateral). This makes the correlation between the radio and CMB bispectrum positive for $\ell \geq 168$. The IR/CMB correlation in turn, becomes positive at slightly higher multipoles, $\ell = 190$, because the IR bispectrum peaks in the squeezed configurations where the CMB bispectrum is still negative. At larger multipoles, because of the complex pattern of the CMB bispectrum and the changes of sign with acoustic peaks, the correlation coefficients tend asymptotically to zero.\\
The IR and radio bispectrum templates are positively correlated as they are both positive ; this correlation is small when analysing only the lowest multipoles, but it increases rapidly over 80\% with a small 'dip' at $\ell_\mathrm{max}=200-300$ at the location of the first CMB acoustic peak. Indeed the CMB spectrum is involved in the denominator of Eq.\ref{Eq:defscalprod}, so at $\ell_\mathrm{max}=200-300$ it typically down-weights non-squeezed configurations where radio and IR bispectra are the most correlated. Nevertheless, the correlation between radio and IR bispectra asymptotically tends to 1 at high angular resolutions, as the IR bispectrum flattens at high multipoles.

\begin{figure}[htbp]
\begin{center}
\includegraphics[width=1.\linewidth]{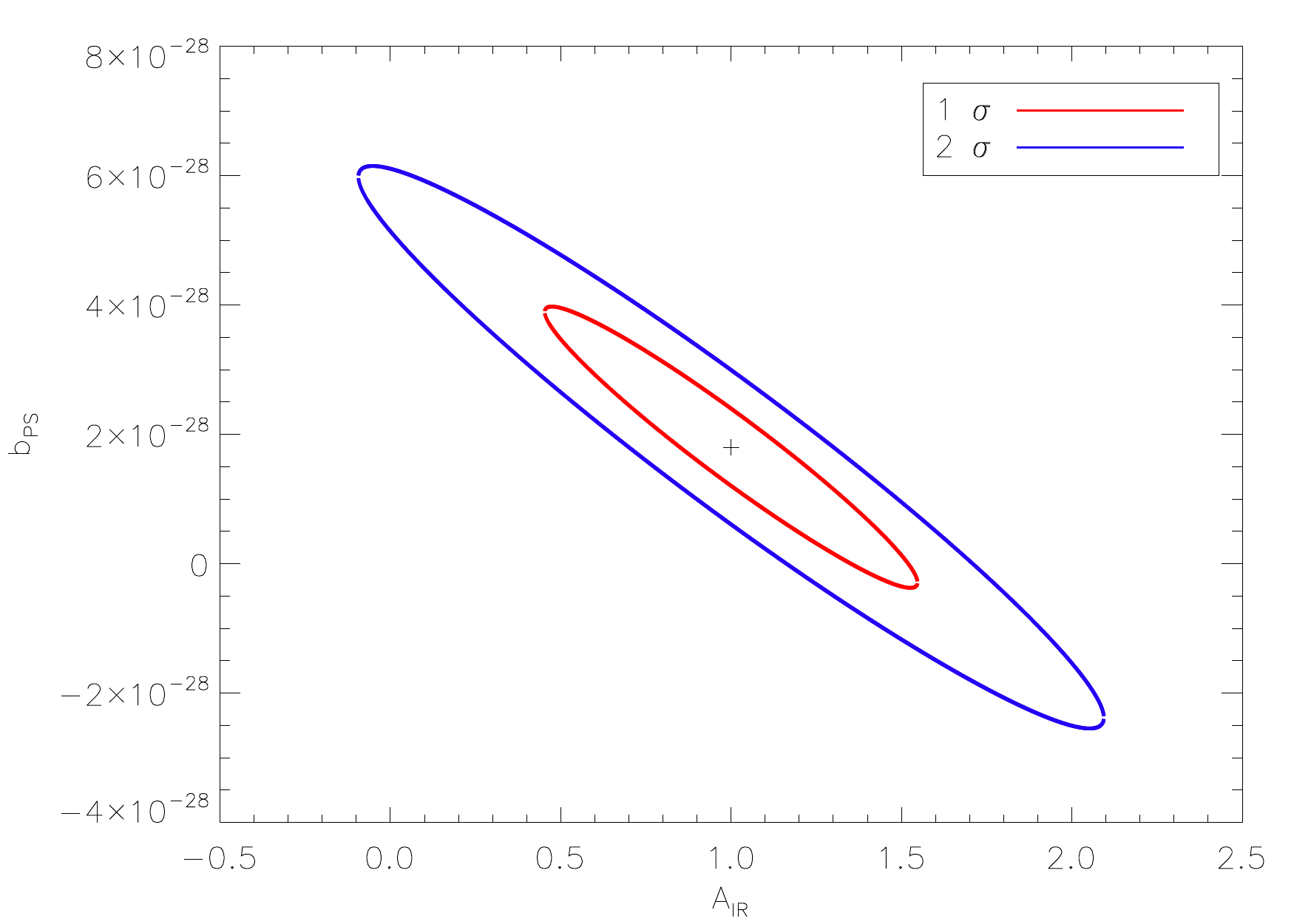}
\caption{1$\sigma$ and 2$\sigma$ confidence level for ($\air$ , $\bps$) in a joint estimation at 220 GHz $\ell_\mathrm{max} = 2048$ with 5 arcmin Gaussian beam and $10^{-8}\, \Delta T/T$ noise}
\label{Fig:ellipse}
\end{center}
\end{figure}

Finally, in Fig.\ref{Fig:ellipse} we forecast the likelihood contours in the ($\air$ , $\bps$) plane, that we expect with the aforementioned instrumental specifications.
$\air$ and $\bps$ estimations are quite degenerate, as expected from the high correlations of their templates, which significantly degrades the constrain that can be put on each of them independently. However we can still have an order of magnitude estimation for both of them. Also, including the prior $\bps \geq 0$ (and also to some extent $\air \geq 0$) would improve the constraints. Finally, CMB removal through component separation may also help by decreasing the variance of both estimators (although not their correlation).

%*********************************************************************

\section{Conclusions and discussion} \label{Sect:concl}

The prescription proposed in \cite{Lacasa2012} yields a separable form for the bispectrum of IR point-sources. This permits us to build a fast estimator for its amplitude \`{a}-la-KSW \citep{KSW2005}. We show how this estimator can be optimised to account for partial sky coverage and inhomogeneous noise. We compute the detection significance that can be expected at several frequencies.  
 Finally we show how this estimator can be used together with estimators of the radio and CMB non-Gaussianity to build a joint estimation of sources of non-Gaussianity. We find that the estimation of CMB NG is barely correlated with the estimation of point-sources NG, so that unless the latter is too important --e.g. if few sources are masked--, both estimations can be considered decoupled. On the contrary, for point-sources, IR and radio NG have similar shapes especially at high multipoles which make their amplitude estimation degenerate. We show how this hampers the detectability of the IR NG at 220 GHz. Increasing this detectability can be achieved for example by subtraction of the CMB, or by including priors on $\bps$ in the analysis (e.g. positivity and/or upper limits given the flux cut and radio source models).

This study opens up the possibility of robust measurements of non-Gaussianity where contaminations are treated in a systematic way. It also opens up the possibility of the detection of point-sources NG, which may permit to constrain their respective models.

%********************************************************************

\section*{Acknowledgments}
The authors thank B. Wandelt for interesting discussions. They acknowledge the use of the \verb"HEALpix" package \citep{Gorski2005}

%**********************************************************************

%\appendix
%\section{Appendix}

%***********************************************************************************

\bibliographystyle{aa}
\bibliography{article}

%***********************************************************************************

\end{document}